\documentclass[aps,pre,twocolumn,showpacs,superscriptaddress,longbibliography]{revtex4-1}
\usepackage{graphicx}
\usepackage{amsmath}
\usepackage{amssymb}
\usepackage[pdftex,bookmarks,colorlinks,breaklinks]{hyperref}  
\hypersetup{linkcolor=red,citecolor=blue,filecolor=dullmagenta,urlcolor=blue} 
\hypersetup{
 pdfauthor={K. Oubaha, M. Richter, U. Kuhl, F. Mortessagne, O. Legrand},
 pdftitle={Refining the Experimental Extraction
  of the Number of Independent Samples
  in a Mode-Stirred Reverberation Chamber},
 pdfkeywords={}, pdfsubject={}, pdfcreator={}, pdflang={English} }

\usepackage{amsmath,amssymb,amsfonts}
\usepackage{algorithmic}
\usepackage{graphicx}
\usepackage{textcomp}
\usepackage{color}
\usepackage{grffile}   
\usepackage{multirow}

\begin{document}

\title{Refining the Experimental Extraction\\
  of the Number of Independent Samples\\
  in a Mode-Stirred Reverberation Chamber\\
}

\author{Khalid Oubaha}
\author{Martin Richter}
\author{Ulrich Kuhl}
\author{Fabrice Mortessagne}
\author{Olivier Legrand}
\affiliation{
Institut de Physique de Nice, Universit\'{e} C\^{o}te d'Azur, CNRS, 06100 Nice, France}

\providecommand*{\Nf}{\ensuremath{N_f}}
\providecommand*{\stirr}{\ensuremath{\theta}}
\providecommand*{\Nstirr}{\ensuremath{N_\theta}}
\providecommand*{\ue}{\ensuremath{\mathrm{e}}}
\providecommand*{\ui}{\ensuremath{\mathrm{i}}}
\providecommand*{\ud}{\ensuremath{\mathrm{d}}}
\providecommand*{\Nind}{\ensuremath{N}}
\providecommand*{\Nindfit}{\ensuremath{N^{*}}}
\providecommand*{\corrstirr}{\ensuremath{\lambda_{\stirr}}}
\providecommand*{\corrstirrnorm}{\ensuremath{\lambda_{\stirr}}}
\providecommand*{\corrstirrfit}{\ensuremath{\lambda^{*}_{\stirr}}}
\providecommand*{\TA}{\ensuremath{T_\mathrm{A}}}
\providecommand*{\TB}{\ensuremath{T_\mathrm{B}}}
\providecommand*{\Deltaf}{\ensuremath{\Delta_{f}}}
\providecommand*{\Deltastirr}{\ensuremath{\Delta\stirr}}
\providecommand*{\Tampl}{\ensuremath{S_{21}}}
\providecommand*{\Trans}{\ensuremath{|\Tampl|^2}}
\providecommand*{\fmax}{\ensuremath{f_\mathrm{max}}}
\providecommand*{\deltafmax}{\ensuremath{\delta\fmax}}
\providecommand*{\LUF}{\ensuremath{f_\mathrm{LUF}}}
\providecommand*{\NIS}{\ensuremath{\mathrm{NIS}}}
\providecommand*{\CRC}{\ensuremath{\mathrm{CRC}}}
\providecommand*{\Vstirr}{\ensuremath{V_\mathrm{stirrer}}}
\providecommand*{\Nsmall}{\ensuremath{\Nind_\mathrm{s}}}
\providecommand*{\Nlarge}{\ensuremath{\Nind_\mathrm{l}}}
\providecommand*{\Csmall}{\ensuremath{C_\mathrm{s}}}
\providecommand*{\Clarge}{\ensuremath{C_\mathrm{l}}}
\providecommand*{\Nindhallbj}{\ensuremath{\Nind^\mathrm{interpol}}}
\providecommand*{\fixme}[1]{\textsc{\textcolor{red}{#1}}}
\providecommand*{\multiplecorr}{\ensuremath{\mathrm{3}}}
\providecommand*{\multipledeltaf}{\ensuremath{\mathrm{8}}}

\date{May 16, 2018}

\begin{abstract}
We investigate the number of independent samples in a chaotic
reverberation chamber.
Its evaluation as defined by the IEC standard can be
made more precise when using not the index of the first
value larger then the correlation length
but using the value obtained by a linear interpolation instead.
The results are validated by a juxtaposition with values from a
measurement using a high stirrer-angle resolution.
A comparison with estimates known from the literature validates
our findings.
An alternative approach using the local maxima of the
parametric dependence of the transmission is
presented in order to show the applicability of the extracted
correlation length over a large range of frequencies.
\end{abstract}

\maketitle

\section{Introduction}

Mode-stirred reverberation chambers play an important role in
electromagnetic compatibility.
With their help it is possible to obtain statistically valuable results
for the electromagnetic radiation emitted from the object under test.
However, in order to get reliable statements about the fluctuations of
the EM field it is necessary to use statistically independent
experimental realizations.
In other words, the statistical ensemble usually achieved by a so-called
mode stirrer has to be mixing enough to make the intensity patterns of
the chamber statistically independent from each other.
As the solution to Maxwell's equations under the given boundary
conditions depend continuously on the latter, a sufficiently large
change has to be performed.
For mode stirrers, this usually means that the angle of rotation has to
be sufficiently large, assuming the stirrer is not too small.
Once this minimally necessary step width is determined the number of
independent samples (\NIS) from a full turn of the stirrer can be
calculated~\cite{IEC_standard}.
This number is the main focus of this paper.
Note that the results are obtained for an empty chamber.
We do not expect that loading effects of typical devices under test
will have a qualitative impact on our findings in a chaotic
reverberation chamber (\CRC).

Because of being such an important quantity there exist also rough
estimates following geometrical arguments~\cite{hal02}.
They are based on probabilistic estimates for rays of the EM field
hitting the stirrer.
Besides these predictions, one can use experimental data to extract the
\NIS.
As the measured data is available only for a discrete set of stirrer
positions the minimally necessary step width is technically only defined
for these discrete values.
As a consequence the \NIS{} can fluctuate quite strongly with frequency.
This can be overcome by either increasing the resolution of the stirrer
movement or by interpolating the results.
In this paper we will compare this interpolation for measurements
performed in a chaotic reverberation chamber.

\section{Experimental Setup}
\label{sec:exper-setup-transm}

\begin{figure}[tb]
  \centering
  \mbox{
    \includegraphics[width=0.57\linewidth]{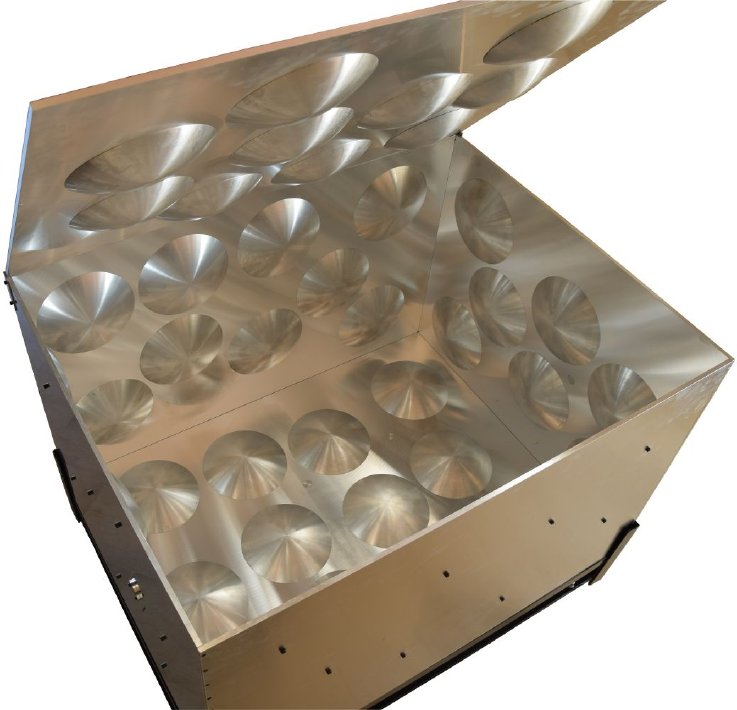}
    \includegraphics[width=0.41\linewidth]{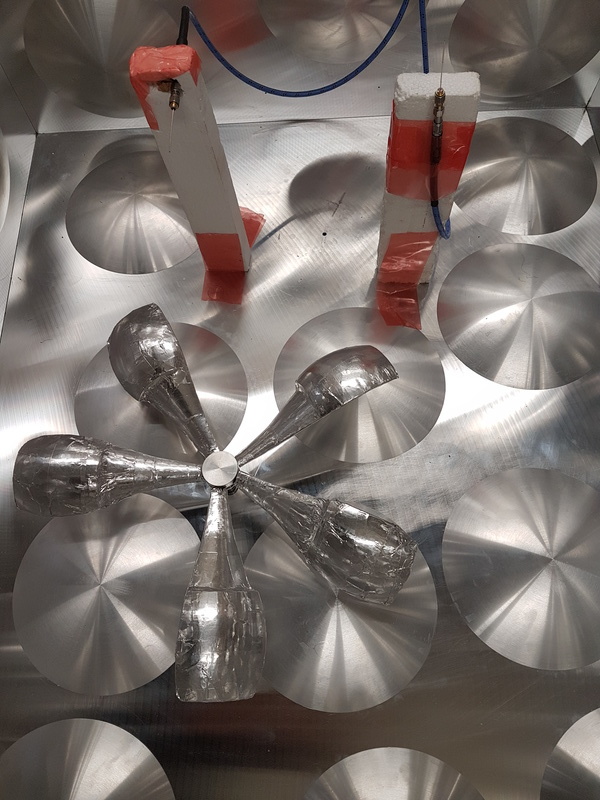}
    }
    \caption{%
      Photograph of the chaotic reverberation chamber with length
      $L = 100\,\mathrm{cm}$, width $W = 77\,\mathrm{cm}$ and height
      $H = 62\,\mathrm{cm}$.
      At the walls $54$ spherical caps of radius $r_\mathrm{c} =
      10\,\mathrm{cm}$ are used, $51$ having a cap height of
      $h_\mathrm{c} = 3\,\mathrm{cm}$ and
      $3$ having $h_\mathrm{c} = 8\,\mathrm{cm}$.
      The total internal volume is $V = 0.44\,\mathrm{m}^3$ (left).
      At the bottom a stirrer with $5$ paddles is placed which can
      be turned by a stepper motor and acts as mode stirrer.
      %
      %
      The two monopole antennas were mounted on
      polystyrene
      blocks having a perpendicular polarization direction (right).
    }
  \label{fig:crc-Nice}
\end{figure}

The experiments were carried out in our chaotic reverberation chamber
shown in Fig.~\ref{fig:crc-Nice}.
The Vector-Network Analyzer (VNA, Rohde\&Schwarz ZVA67) was attached to
two monopole antennas inside the chamber (see
Fig.~\ref{fig:crc-Nice}(right)).
For a total range of frequencies $f$ from $0.5\,\mathrm{GHz}$ to
$5.0\,\mathrm{GHz}$ we measured the complex transmission amplitude
between the two antennas.
The measurement was repeated for $\Nstirr = 3600$ stirrer positions
using a step of $\Deltastirr = 0.1^\circ$,
Eq.~\eqref{eq:angle-resolution}.
This range was chosen to cover frequencies from around the Lowest-Usable
Frequency \LUF~\cite{IEC_standard,gro15,gro16} up to approximately
\(7\cdot\LUF\).
Using the frequency of the $60^\mathrm{th}$
mode as a definition, the
\LUF{} of our homemade chaotic chamber
is approximately at $0.735\,\mathrm{GHz}$.

\begin{figure}[tb]
  \centering
  \includegraphics[width=0.9\linewidth]{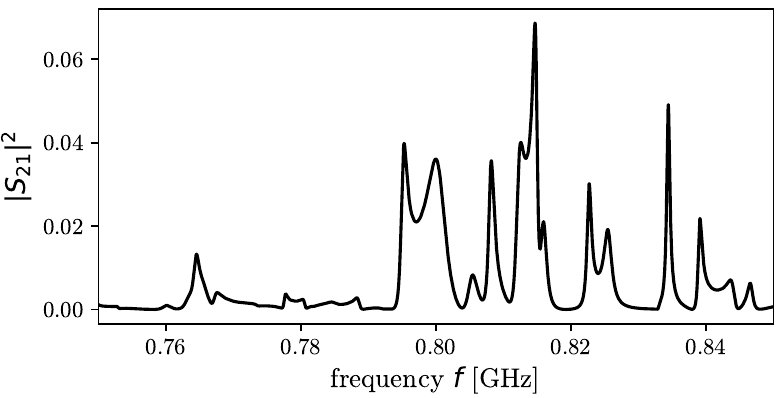}
  \includegraphics[width=0.9\linewidth]{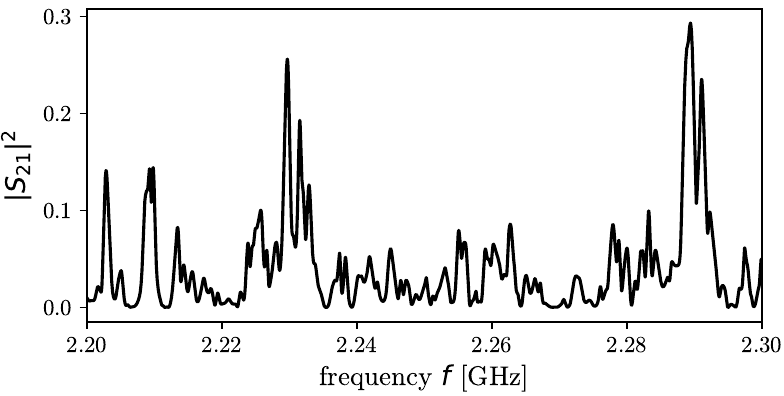}
  \includegraphics[width=0.9\linewidth]{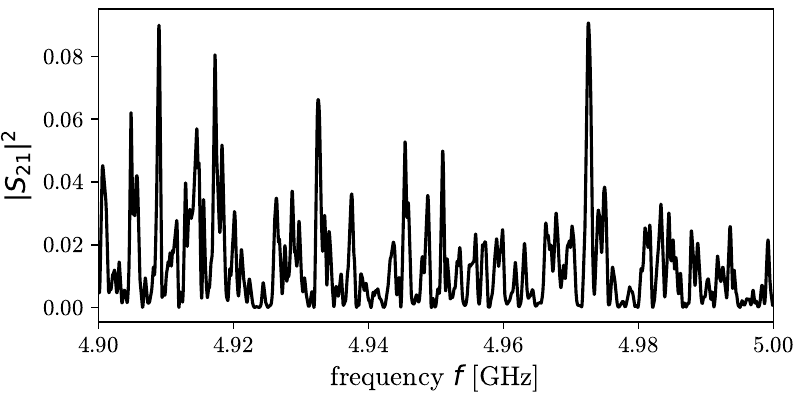}
  \caption{%
    Dependency of the transmission $|\Tampl|^2$ on frequency $f$ for
    one fixed stirrer position $\stirr = 1^\circ$.
    Three frequency ranges,
    representing different regimes of the modal overlap $d$,
    are shown:
    The lowest frequency range (top, $d = 0.3$), the
    highest frequency range (bottom, $d = 4.5$),
    and a frequency range in between corresponding to the
    example in Fig.~\ref{fig:schema-linear-interpolation}
    (middle, $d = 1.4$).
    For the values of $d$ see
    Tab.~\ref{tab:crc-figures-of-merit}.
    The frequency range was chosen to cover
      $100\,\mathrm{MHz}$ in all plots.
  }
  \label{fig:T-freq-dependency-gamma1}
\end{figure}

Corresponding transmissions for three different frequency ranges with
different values of the modal overlap for one fixed stirrer position
versus frequency are shown in Fig.~\ref{fig:T-freq-dependency-gamma1}.

Besides the whole frequency range $0.5 - 5\,\mathrm{GHz}$ we focused on
seven sub-intervals, where we measured with a higher frequency resolution
to guarantee a proper extraction of the transmission maxima.
The ranges have been chosen such that important parameters (see
Tab.~\ref{tab:crc-figures-of-merit}) are sufficiently well defined.
One important parameter is
the mean frequency spacing of adjacent eigenmodes of the cavity,
\(\Deltaf\),
which we calculate using Weyl's law,
$\Deltaf = c^3 / (8\pi V f^2)$.
The signal decay time $\tau$ is determined from the exponential decay of
the square modulus of the Fourier transform of the
transmission~\cite{kuh17b_emc_amsterdam},
\begin{align}
  \label{eq:fft_amplitde_exp_decay}
  I(t)
  &= \left\vert \mathrm{FT}(\Tampl)(t)\right\vert^2
  = I_0\ue^{-t / \tau},
\end{align}
where the transform is
performed either over the frequency range of the seven sub-intervals or a
frequency window of 100\,MHz.
The average quality factor $Q$ is thus given by $Q = 2\pi \tau \langle f \rangle$,
where $\langle f \rangle$ is the average frequency of the window used to
calculate $\tau$.
Finally, the modal overlap $d$ is obtained by
\begin{align}
  \label{eq:modal-overlap}
  d &= \frac{\mathrm{mean\,decay\,rate}}
      {\mathrm{mean\,eigenmode\,spacing}}
      = \frac{1 / 2\pi \tau}{\Deltaf}
      = \frac{\langle f\rangle}{Q\Deltaf}.
\end{align}
The extracted values can be found in Tab.~\ref{tab:crc-figures-of-merit}
for the seven sub-intervals.
\begin{table}[htbp]
  \caption{%
    Figures of merit for the reverberation chamber at different
    frequencies ranges ($f_\mathrm{min} - f_\mathrm{max}$).
    Shown are the values for %
    the mean frequency spacing $\Deltaf$, %
    the signal decay time $\tau$, %
    the mean distance between adjacent maxima $\deltafmax$,
    the quality factor $Q$, %
    and the modal overlap $d$
    The frequency ranges are also indicated in
    Fig.~\ref{fig:Nind-freq_all}.
  }
\begin{center}
  \begin{tabular}{|c@{\,--\,}c|r|r|r|r|r|}
    \hline
    \multicolumn{2}{|c|}{
    $\!\!\frac{f_\mathrm{min} - f_\mathrm{max}}{\mathrm{Ghz}}\!\!$}
    & $\frac{\Deltaf}{\mathrm{kHz}}$
    & $\frac{\tau}{\textrm{ns}}$
    & $\!\frac{\langle \deltafmax \rangle}{\textrm{kHz}}\!$
    & $\frac{Q}{10^{3}}$
    & $d$
    \\
    \hline
    \hline
    0.75 & 0.85 & \!\!3734.43\!\! & 203 & \!\!1422.05\!\! &  0.8 & 0.3 \\ 
    1.1  & 1.25 & \!\!1731.13\!\! & 180 & \!\!1603.75\!\! &  1.0 & 0.7 \\ 
    1.7  & 1.85 & \!\! 758.59\!\! & 203 & \!\!1422.05\!\! &  2.0 & 1.2 \\ 
    2.2  & 2.40 & \!\! 451.80\!\! & 276 & \!\!1045.92\!\! &  3.7 & 1.4 \\ 
    2.8  & 2.90 & \!\! 294.25\!\! & 343 & \!\! 841.62\!\! &  5.9 & 1.6 \\ 
    3.8  & 4.0  & \!\! 157.14\!\! & 345 & \!\! 836.74\!\! &  8.4 & 3.0 \\ 
    4.9  & 5.0  & \!\!  97.54\!\! & 361 & \!\! 799.65\!\! & 11.2 & 4.5 \\ 
    \hline
  \end{tabular}
\label{tab:crc-figures-of-merit}
\end{center}
\end{table}

\section{Refining the Extraction of the Number of Independent Samples}

\begin{figure}
  \centering
  \includegraphics{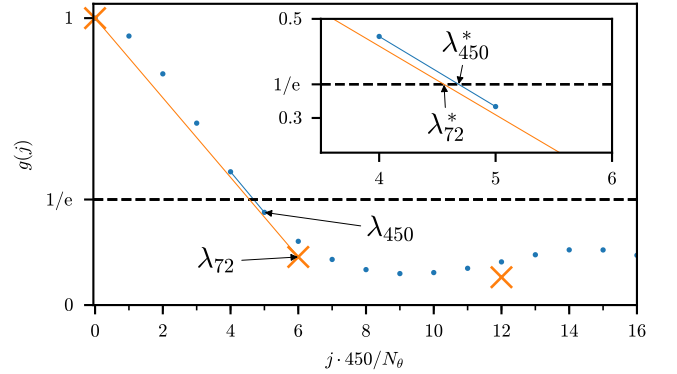}
  \caption{%
    Linear interpolation used to obtain the improved correlation
    length $\corrstirrfit$~\eqref{eq:corr-length-interpolated} from the
    traditional value $\corrstirrnorm$~\eqref{eq:corr-length-integer}.
    The plot is done for $f = 2.21\,\mathrm{{GHz}}$ and therefore
    corresponds to the middle image of
    Fig.~\ref{fig:T-freq-dependency-gamma1}.
    Shown is the extraction for $\Nstirr = 450$ (blue points) and
    $\Nstirr = 72$ (orange crosses).
  }
  \label{fig:schema-linear-interpolation}
\end{figure}

The usual approach to extract the \NIS{} is to use the autocorrelation
function of the transmission data.
Given \Nstirr{} equidistant angles at which the transmission amplitude
$\Tampl(f_i, \stirr_j)$ has been measured,
\begin{align}
  \label{eq:angle-resolution}
  \stirr_j
  &= (j - 1)\cdot \Delta \stirr,
  &
    \Delta \stirr
  &= \frac{360^\circ}{\Nstirr},
  &
    j
  &= 1, \dots, \Nstirr,
\end{align}
we can calculate for a given frequency $f$, see
Refs.~\cite{IEC_standard,lun00,kra05}
\begin{align}
  \label{eq:correlation-function}
  g_f(j) &= \frac{1}{\Nstirr - 1}
           \frac{\sum_{k=1}^{\Nstirr}
           x(f, \stirr_k) x(f, \stirr_{k + j})}
           {\sum_{k=1}^{\Nstirr}x(f, \stirr_{k})^2}
\end{align}
where \(x = \tilde{x} - \langle \tilde{x}(f) \rangle\),
\(\langle \tilde{x}(f) \rangle =
\frac{1}{\Nstirr}\sum_{j=1}^{\Nstirr}\tilde{x}(f, \stirr_j)\), and
\(\tilde{x} = |\Tampl|^2\)
and we use that $\Tampl$ is periodic in \stirr.
In Fig.~\ref{fig:schema-linear-interpolation} an example from our
measured data is shown for $\Nstirr = 450$ (blue points) and
$\Nstirr = 72$ (orange crosses).
For simplicity we only use one stirrer although generalizations to
multiple stirrers exist~\cite{gra13}.

Assuming an exponential decay of correlations,
$g_f(j) \sim \ue^{-j / \lambda(f)} $, one defines the correlation length
\corrstirrnorm{} as the smallest integer
for which
\begin{align}
  \label{eq:corr-length-integer}
  g_f(\corrstirrnorm(f)) \le \ue^{-1} \approx 0.37.
\end{align}
In case of small sample size the
definition has to be refined by using (see Eq.~(A.5) in
Ref.~\cite{IEC_standard})
\begin{align}
  \label{eq:corr-length-integer-refined}
  g_f(\corrstirrnorm(f)) \le 0.37
  \left(1 - \frac{7.22}{\Nstirr^{0.64}}\right).
\end{align}
Note that Eq.~\eqref{eq:corr-length-integer} was given in the IEC
standard version of 2003. The refinement
Eq.~\eqref{eq:corr-length-integer-refined} was suggested in the version
of 2011.
While this alters the determined \NIS, the
change is not of qualitative nature
as can be seen from the comparison of
Eq.~\eqref{eq:corr-length-integer} and
Eq.~\eqref{eq:corr-length-integer-refined} in
Fig.~\ref{fig:Nind-freq_all}.
The main difference is seen as expected for small values like
\(\Nstirr = 72\) which, for the older norm saturates at its maximum
value $N = 72$ for large frequencies.
For the current norm this saturation is not reached.
In Fig.~\ref{fig:Nind-freq_all} (lower) we apply the current norm also
for $\Nstirr = 72$ although it is strictly speaking just valid for
$\Nstirr \ge 100$.
A detailed analysis of the literature on which the norm is
based~\cite{kra07_for_amsterdam} reveals that there is no
apparent reason for not using the norm for $\Nstirr = 72$.
In the following we use the older norm,
Eq.~\eqref{eq:corr-length-integer}, as it makes the comparison for
different sample sizes \Nstirr{} easier to understand.
The \corrstirrnorm{} are indicated by $\lambda_{450}$ and $\lambda_{72}$
in Fig.~\ref{fig:schema-linear-interpolation}.
This value is usually used in the literature to define the number of
independent field components as~\cite{IEC_standard,hal02}
\begin{align}
  \label{eq:N-independent-samples}
  \Nind(f) = \frac{\Nstirr}{\corrstirrnorm(f)}.
\end{align}

If the correlation decays quickly, then the extracted value of
$\corrstirrnorm(f)$ fluctuates between small integer values when
changing the frequency.
Therefore, \Nind{} might express large fluctuations.
These would vanish if a smaller $\Deltastirr$ would be used as this
would result in a higher resolution of \corrstirrnorm{} according to
Eq.~\eqref{eq:corr-length-integer}, see Fig.~\ref{fig:Nind-freq_range7}
for example.

A refined method can be obtained by linearly interpolating
the two points before and after the critical
value~\eqref{eq:corr-length-integer} is undercut,
\begin{align}
  \label{eq:corr-length-interpolated}
  \corrstirrfit
  & = \corrstirrnorm -
    \frac{\frac{1}{\ue} - g(\corrstirrnorm)}{
    g(\corrstirrnorm - 1) - g(\corrstirrnorm)}
  \\
  \label{eq:Nind-interpolated}
  \Nindfit(f)
  &= \frac{\Nstirr}{\corrstirrfit(f)}.
\end{align}
This approach is justified by the fact that a larger \Nstirr{} and
therefore higher angle
resolution~\eqref{eq:angle-resolution} yields a finer resolution of the
autocorrelation function.
In the inset of Fig.~\ref{fig:schema-linear-interpolation} the values
obtained by using this fit for two different \Nstirr{} are shown.
The resulting values are quite close, whereas using the index directly
would lead to large differences in \Nind.
In our experiment we checked this assumption by comparing the data with
a reduced data set like in Ref.~\cite{kra05}, see
Sec.~\ref{sec:exper-setup-transm}.
Note that due to the fact that
$\corrstirrnorm - 1 < \corrstirrfit \le \corrstirrnorm$, the
extracted \NIS{} fulfills $\Nindfit \ge \Nind$.
Furthermore, using the decay of the autocorrelation function only
estimates the number of uncorrelated samples.
However, this approach has become normative for the determination of
the \NIS{} in the literature.

\subsection{Number of Independent Samples}

We calculated the number of independent
samples~\eqref{eq:N-independent-samples} based on the integer-valued
correlation length~\eqref{eq:corr-length-integer} and the interpolated
value~\eqref{eq:Nind-interpolated} based on
Eq.~\eqref{eq:corr-length-interpolated}, respectively.
The analysis was done once for the whole frequency range
$0.5 - 5\,\mathrm{GHz}$.
We determine the independent samples using either the whole set of
measured angles ($\Nstirr = 3600$, $\Deltastirr = 0.1^\circ)$ or two
reduced data sets~\cite{kra05} with $\Nstirr = 450$
$(\Deltastirr = 0.8^\circ)$ and $\Nstirr = 72$
$(\Deltastirr = 5^\circ)$.
The value of $\Nstirr = 450$ is commonly used in the literature as
suggested in the standard~\cite{IEC_standard} but not without
criticism~\cite{lun00,kra05,lem07}.

In order to check whether the linear interpolation of
formula~\eqref{eq:corr-length-interpolated} works for our experimental
data, it is shown in Fig.~\ref{fig:schema-linear-interpolation} for
$f = 2.21\,\mathrm{{GHz}}$ using the \(\Nstirr = 450\) (shown as blue
dots) and for a strongly reduced number of \(\Nstirr = 72\) (shown as
large crosses).
On the one hand one can see that the values $\lambda_{450}$ and
$\lambda_{72}$ obtained using the discrete numbers deviate, whereas the
values of $\lambda^*_{450}$ and $\lambda^*_{72}$ are quite close (after
an appropriate rescaling).
In this example one can also see that the interpolation is justified as
the decay of the correlation around \(1 / \ue\) is approximately linear.
Have in mind that Fig.~\ref{fig:schema-linear-interpolation}
is using a rescaled value of the integer $j\cdot 450/\Nstirr$ due to the
reduction of angles, so that the axis corresponds to the index $j$ in
the case $\Nstirr=450$.
Note that the interpolation also works close to \LUF{} as our
reverberation chamber is rendered fully chaotic due to the spherical
caps at the walls~\cite{gro15}.

\begin{figure}[tb]
  \centering
    \includegraphics[width=1.0\linewidth]{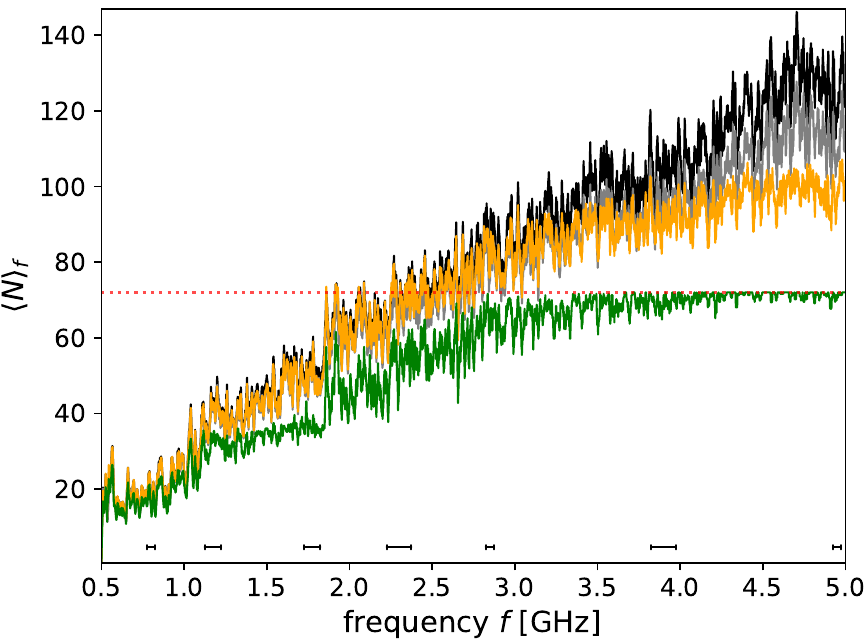}
    \\
    \includegraphics[width=1.0\linewidth]{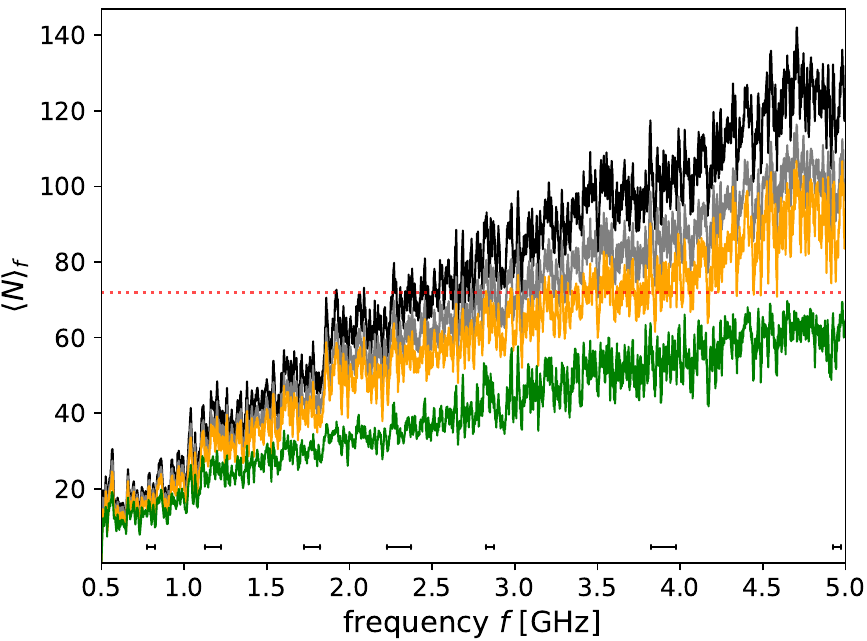}
    \caption{%
      Number of independent samples for frequency range
      $0.5-5\,\mathrm{GHz}$ determined by the minimal
      integer~\eqref{eq:N-independent-samples} and by the
      linear interpolation~\eqref{eq:Nind-interpolated}
      (upper) as well as the curves given by
      Eq.~\eqref{eq:N-independent-samples} based on a correlation
      length from Eq.~\eqref{eq:corr-length-integer-refined}
      (lower).
      The different curves belong to the different angle resolutions,
      namely
      $\Nstirr = 3600$ (black),
      $\Nstirr = 450$ (gray),
      $\Nstirr = 72$ (green).
      For $\Nstirr = 72$ the values obtained by linear
      interpolation~\eqref{eq:Nind-interpolated} are also shown
      (orange).
      The black intervals at the bottom indicate the ranges used in
      Tab.~\ref{tab:crc-figures-of-merit}.
      The red dotted line indicates $\Nind = 72$.
      All curves show frequency-smoothed values after applying a
      rectangular-filter of $100$ frequency steps ($10\,\mathrm{MHz}$).
    }
  \label{fig:Nind-freq_all}
\end{figure}
The overall dependency of the \NIS{} versus the full frequency range for
different calculation methods is shown in Fig.~\ref{fig:Nind-freq_all}.
Because one observes large fluctuations with frequency, we applied a
rectangular frequency filter.
In general, such large fluctuations make the extraction of the \NIS{} by
using only a single frequency questionable.
For increasing frequencies the averaged \NIS{} rises as the stirrer
position is better resolved by the EM field in accordance with the
estimates from Ref.~\cite{hal02}.
This higher resolution of the stirred volume increases the sensitivity
with respect to the stirrer position and thereby decreases the
correlation length.
The data using the 3600 (black curve) and 450 (orange curve) samples are
following each other closely apart from small deviation at higher
frequencies.
In case of 72 samples the \NIS{} is bounded at \(\Nind = \Nstirr\) (red
dotted line).
In this case, \(\corrstirrnorm\) is always $1$ as the first
corresponding value of $g_f$ in Eq.~\eqref{eq:corr-length-integer}
already lies below the threshold, $g_f(1) \le 1 / \ue$ (see also
Fig.~\ref{fig:schema-linear-interpolation}).
Additionally we plotted the curve for 72 samples where we used the
linear interpolation to obtain $\lambda^*_{72}$ to calculate the \NIS{}
(see Fig.~\ref{fig:Nind-freq_all}, orange curve).
This curve follows better the curves with larger \Nstirr{} and even
gives values above $72$.
While technically the $72$ sample cannot have more than $72$ independent
data sets, the linear interpolation outlines a possibility to estimate
which angular resolution is a good choice for getting as many
independent samples with a minimal amount of measurements.

Indicated in the figure are also the seven frequency sub-ranges, for which
parameters are detailed in Tab.~\ref{tab:crc-figures-of-merit}.
\begin{figure}[tb]
  \centering
  \includegraphics[width=1.0\linewidth]{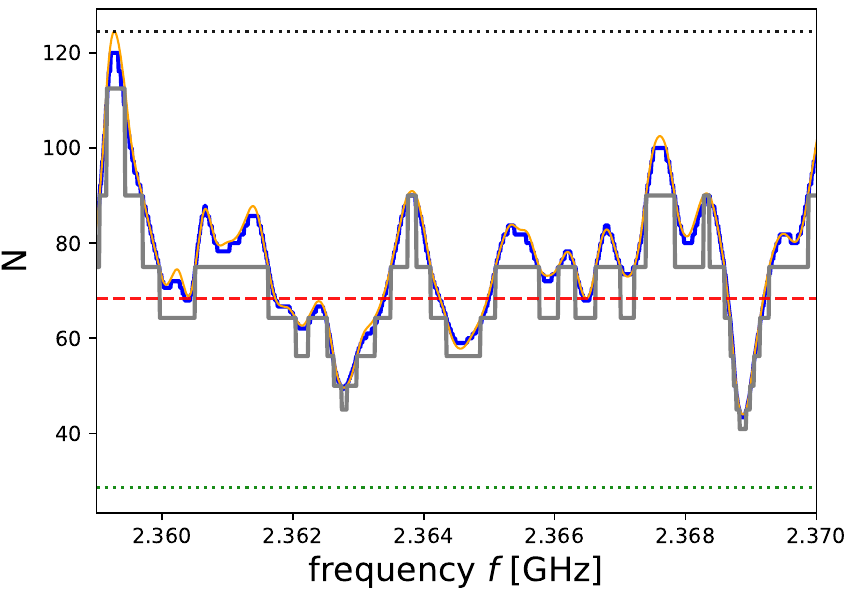}
  \includegraphics[width=1.0\linewidth]{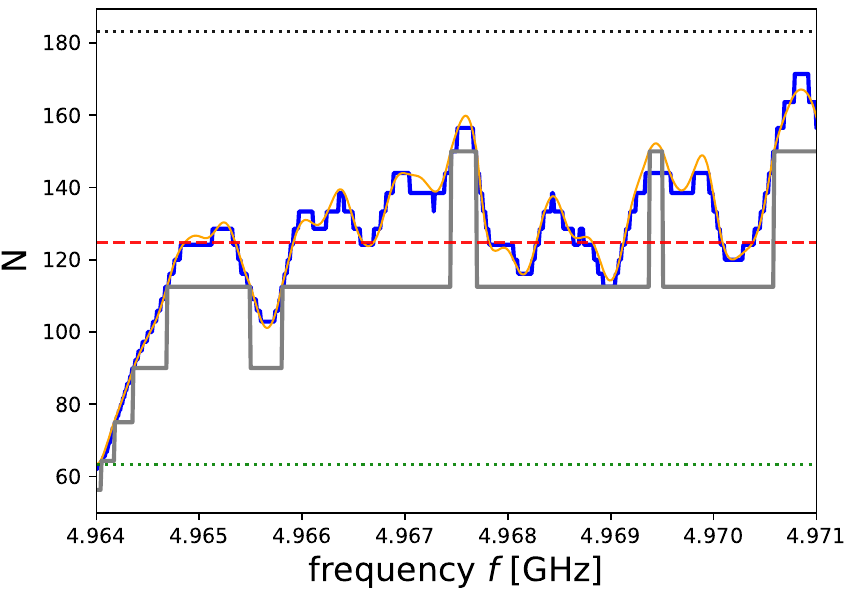}
  \caption{
    Number of independent samples for two
    small frequency ranges.
    Shown is a zoom of the data from two subsets of
    Tab.~\ref{tab:crc-figures-of-merit}
    ($2.2-2.4\,\mathrm{GHz}$ top,
    $4.9-5\,\mathrm{GHz}$ bottom).
    We compare the standard approach
    Eq.~\eqref{eq:N-independent-samples} for $\Nstirr = 450$ (gray) and
    $\Nstirr = 3600$ (blue) with \Nindfit{} obtained for  $\Nstirr =
    450$ (orange) via the interpolation
    Eq.~\eqref{eq:Nind-interpolated}.
    The maximum and minimum values (black and green dotted lines)
    as well as the average (red dashed line) of the interpolation
    for $\Nstirr = 450$ (orange curve) over
    the whole corresponding frequency range of the corresponding
    sub-window (see Tab.~\ref{tab:crc-figures-of-merit}) are added for
    comparison.
    In contrast to Fig.~\ref{fig:Nind-freq_all}, no frequency average is
    applied.
  }
  \label{fig:Nind-freq_range7}
\end{figure}
Fig.~\ref{fig:Nind-freq_range7} shows $\Nind(f)$ and $\Nindfit(f)$ for
two of these frequency sub-sets
as well as for different stirrer resolutions $\Nstirr$.
Here, no frequency average was applied.
Note that the smaller sample size leads to larger step sizes in the
frequency axis in case of the calculation via
Eq.~\eqref{eq:N-independent-samples}.
On the one hand side we find that the larger $\Nstirr=3600$ is always
above the smaller $\Nstirr=450$ value but the value obtained by the
linear interpolation for $\Nstirr=450$ follows nicely the $\Nstirr=3600$
case.

\subsection{Prediction of the Number of Independent Samples}

\begin{figure}[tb]
  \centering
  \includegraphics[width=0.9\linewidth]{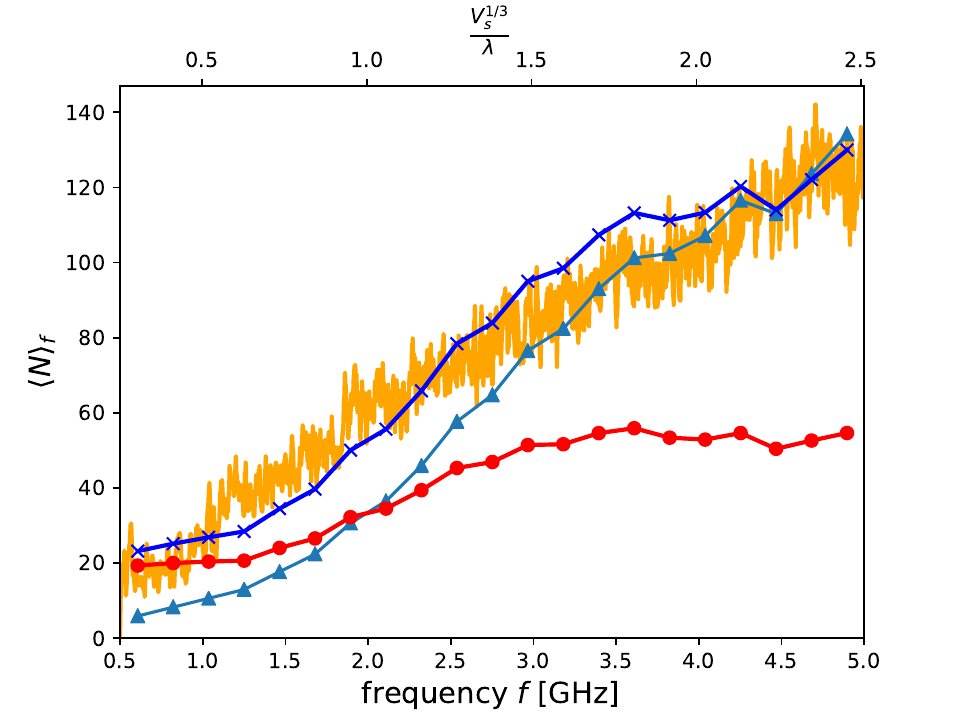}
  \caption{%
    Comparison of the number of independent samples with the
    estimate based on geometrical arguments.
    Based on the experimental data for $\Nstirr = 3600$ we
    fit the prefactor in Eq.~\eqref{eq:hallbjoerner-small} (red circles)
    and Eq.~\eqref{eq:hallbjoerner-large} (light blue triangles).
    The interpolating formula~\eqref{eq:hallbjoerner-interpolation}
    matches the whole frequency range (dark blue curve).
    The corresponding values are
    $\Csmall = 1.87$, $\Clarge = 0.7$.
    The experimental curve show frequency-smoothed values after
    applying a rectangular-filter of $100$
    frequency steps ($10\,\mathrm{MHz}$).
  }
  \label{fig:comparison-estimate-Vstirrer}
\end{figure}

The above extracted values for the number of independent samples can be
compared with an estimate based on the volume $V$ and quality factor $Q$
of the chamber~\cite{hal02}.
The prediction is based on a probabilistic argument that the
stirred volume \Vstirr{} is hit by a beam in the chamber.
It differs for large and small stirrers, \Nlarge{} and \Nsmall,
respectively.
In our case the volume affected by the stirrer as defined in
Ref.~\cite{hal02} is
\begin{align}
  \label{eq:V_stirrer}
  \Vstirr = 0.0034\,\mathrm{m}^3.
\end{align}
For the low frequency range the estimate is given by~\cite{hal02}
\begin{align}
  \label{eq:hallbjoerner-small}
  \Nsmall
  & =\Csmall \frac{\lambda \Vstirr^{2/3}}{V}Q
    ,
  & \Vstirr \ll \lambda^3
\end{align}
whereas in the high frequency range it should be related to
\begin{align}
  \label{eq:hallbjoerner-large}
  \Nlarge
  &=\Clarge \frac{\Vstirr}{V}Q
    ,
  & \Vstirr \gg \lambda^3.
\end{align}
When both expressions are adjusted to the experimental data in the
appropriate frequency range we obtain $\Csmall = 1.55$ and
$\Clarge = 1.56$, respectively.
In Fig.~\ref{fig:comparison-estimate-Vstirrer} the averaged number of
independent samples is compared to the two predictions.
A good agreement is found to \eqref{eq:hallbjoerner-small} (red circles)
in the small frequency range ($\Vstirr^{1/3}/\lambda <0.5$, see upper
axis) and to \eqref{eq:hallbjoerner-large} (blue triangles) in the high
frequency range ($\Vstirr^{1/3}/\lambda >1.75$, see upper axis).
Also the values of \Csmall{} and \Clarge{} are of the order of one
agreeing with values obtained in \cite{hal02}.
In view of the two curves we propose here an interpolating estimate with
two fitted prefactors which are compatible with the extracted values
over the full frequency range.
This formula reads
\begin{align}
  \label{eq:hallbjoerner-interpolation}
  \Nindhallbj(f) = \left(\Clarge +
  \Csmall\frac{c / f}{V^{1/3}_\mathrm{stirrer}}\right)
  \frac{\Vstirr}{V}Q(f)
\end{align}
where we use the frequency dependent values of the $Q$ factor
$Q = 2\pi \tau \langle f \rangle$
to fit
Eq.~\eqref{eq:hallbjoerner-interpolation} against the data.
The fit using only small and large frequency values yields
$\Csmall = 1.87$, $\Clarge = 0.7$.
The resulting dark blue curve is shown in
Fig.~\ref{fig:comparison-estimate-Vstirrer}.
The estimate \eqref{eq:hallbjoerner-interpolation} follows reasonably
well over the whole frequency range.

\subsection{Velocities of Local Maxima}

\begin{figure}[htbp]
  \center
  \includegraphics[width=0.95\linewidth]{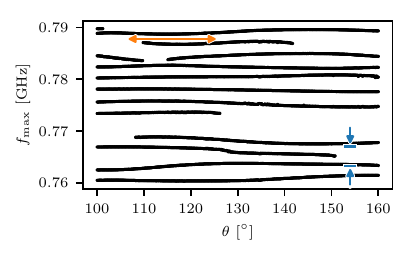}
  \\[-3ex]
  \includegraphics[width=0.95\linewidth]{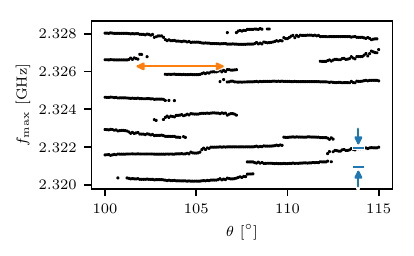}
  \\[-3ex]
  \includegraphics[width=0.95\linewidth]{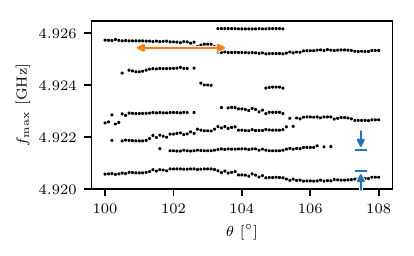}
\caption{\label{fig:fmax}
  Dependency of the local maxima \fmax{} on stirrer position \stirr.
  Depicted are three of the seven frequency ranges from
  Tab.~\ref{tab:crc-figures-of-merit}.
  The horizontal arrow indicates the correlation length \corrstirr{}.
  The vertical arrows indicate \Deltaf{} (upper figure) or \deltafmax{}
  (middle and lower figure).
  The abscissa are scaled to cover $\multiplecorr\cdot\corrstirr$.
  The ordinates are scaled to cover $\multipledeltaf\cdot\Deltaf$ (upper
  figure) or $\multipledeltaf\cdot\deltafmax$ (middle and lower figure).
}
\end{figure}

We can extract the local maxima \fmax~\cite{kuh17b_emc_amsterdam} from
the measured transmission in the seven sub-intervals mentioned in
Tab.~\ref{tab:crc-figures-of-merit}.
In order to get reliable results we smoothed the $|\Tampl(f, \stirr)|^2$
along the frequency axis using a Hann filter of window size $100$ for
all data sets.
We then extracted the local maxima of the intensity for each value of
the stirrer position.
In the following we want to demonstrate that the statistical properties
of the maxima are similar to one another across the frequency ranges
shown, once the appropriate scaling has been applied.
If the resonances are isolated (i.e., the modal overlap $d$ is small
($d\ll 1$)) the local maxima are defined by the frequencies of the
eigenmodes $\nu_n$.
The dependence of the eigenvalues on a parameter $p$ have been studied
extensively in the framework of ``Quantum Chaos'' \cite{haa01b_emc_amsterdam,stoe99_emc_amsterdam}
for scalar fields, but can be directly applied to the vectorial problem.
Defining a level velocity $v_n=\ud\nu_n/\ud p$ one can distinguish
global and local perturbation.
In case of a global perturbation in a \CRC
the distribution of the level
velocities $v_n$ is Gaussian\cite{sim93a}, whereas for local
perturbations it shows a Bessel $K$ distribution \cite{bar99d}.
This would be another possibility to characterize the quality of the
stirring.
In case of a chaotic system the levels show avoided crossing, whereas in
case of regular, more precisely, integrable systems, the levels will
cross\cite{haa01b_emc_amsterdam,stoe99_emc_amsterdam}.
The frequency scale of importance is the mean frequency spacing \Deltaf.
In Fig.~\ref{fig:fmax}(top) the dynamics of the local maxima is
presented for the low frequency range, which has $d=0.3$, thus showing
reasonably isolated resonance.
The spectra should be uncorrelated when the local maxima go from one
avoided crossing to another, which agrees visually with the calculated
correlation length \corrstirr{} indicated by the horizontal arrow.
Studies also exist in the case of open systems \cite{bul06} on eigenmode
dynamics, but in the case of moderate or large modal overlap, the
frequencies of the eigenmodes are not directly related to the local
maxima we extracted here.
To define the frequency scale in case of strong modal overlap we need to
estimate the mean spacing between transmission maxima, which has been
obtained by Schroeder and Kuttruff~\cite{sch62a}, 
$\deltafmax = 1 / (2\sqrt{3}\tau)$,
an expression which is assumed to be valid for $d > 3$.
If abscissa and ordinate are scaled appropriately the behavior of the
ridges of maxima $\fmax(\stirr)$ is similar with respect to the
distances of close encounters and the steepness with respect to the
stirrer angle, i.e., $\frac{\ud\fmax}{\ud\stirr}$.
Using the correlation length \corrstirr\ for each of these frequency
ranges as read from, e.g., Fig.~\ref{fig:Nind-freq_range7}, we can
choose the scale of the abscissa of these plots to cover several
correlation lengths.
We chose the \stirr{} range from \(\stirr_\mathrm{min} = 100^\circ\) to
$\stirr_\mathrm{max}
= \stirr_\mathrm{min} + \multiplecorr \cdot
\langle \corrstirrfit(f) \rangle_{f}
\cdot\Deltastirr$
to cover \multiplecorr{} times the correlation length in each figure.
The ordinate was scaled using the values from
Tab.~\ref{tab:crc-figures-of-merit} in the following way:
For frequency ranges with a modal overlap smaller than $d \le 1$ we
chose a plot range of width $\multipledeltaf \cdot \Deltaf$.
For the other ranges we used the
estimate~$\deltafmax$
and chose $\multipledeltaf \cdot \deltafmax$.
The corresponding plots are shown in Fig.~\ref{fig:fmax} for three of
the seven intervals from Tab.~\ref{tab:crc-figures-of-merit}.
Each of the plots contains two arrows indicating the correlation length
\corrstirr{} and the mean spacing between maxima, respectively.
Due to the scaling of the abscissa the one for \corrstirr{} has the same
length in every plot.

We can indeed see a qualitative agreement between the average distance
between the extracted \fmax{} ridges, thus confirming the results
obtained in the previous chapter.

\section{Conclusion}

This paper presents experimental data from a mode-stirred chaotic
reverberation chamber and estimates the number of independent samples
(\NIS) for a $360^\circ$ turn of the stirrer.
We compare the reduced data set for $\Nstirr = 450$ steps usually found
in the literature with a finer ($\Nstirr = 3600$) and
coarser subdivision ($\Nstirr = 72$) of the full angle.
The corresponding values of the \NIS{} extracted by the usual
procedure~\eqref{eq:N-independent-samples} show a very coarse
dependency if \Nstirr{} is reduced.
We compare this value with an improved estimate based on a linear
interpolation~\eqref{eq:corr-length-interpolated} of the correlation
length.
The values obtained in this way for the coarser measurements resemble
very much the full data set as shown, for example, in
Fig.~\ref{fig:Nind-freq_range7}.
We compare these numbers also with
predictions~\eqref{eq:hallbjoerner-small},~\eqref{eq:hallbjoerner-large}
from Ref.~\cite{hal02} and find quantitative agreement if we
use the interpolation formula~\eqref{eq:hallbjoerner-interpolation}, see
Fig.~\ref{fig:comparison-estimate-Vstirrer}.
We check that the results are valid for conceptually different regimes,
i.e.\ frequency ranges for which the modal overlap $d$ of the chamber is
smaller, around, or larger than $1$, see
Tab.~\ref{tab:crc-figures-of-merit}.

To emphasize the findings with an independent approach, we focus on the
extraction of local maxima of the
transmission~\cite{kuh17b_emc_amsterdam}.
The detailed resolution of $\Nstirr = 3600$ allows to follow
$\fmax(\stirr)$ as the stirrer moves.
The corresponding dynamics show fluctuations on a scale which is
expected to be similar to the correlation length \corrstirr.
Hence, we show that scaling a plot of \fmax{} using \corrstirr{}
yields qualitatively the same image.

\section*{Acknowledgment}

We would like to thank the European Commission for financial support
through the H2020 programme by the Open Future Emerging Technology
``NEMF21'' Project (664828).

\end{document}